\documentclass[
superscriptaddress,
preprint,
amsmath,amssymb,
aps,
floatfix,
showkeys,
]{revtex4-2}

\usepackage{setspace}
\usepackage{amsfonts}
\usepackage{graphicx}
\usepackage{siunitx}
\usepackage{hyperref}
\usepackage{xcolor}

\begin{document}

\setstretch{1.0}

\title{Emergent Spin-Singlet Pairing in the Frustrated Kagome Metal Sc$_3$Mn$_3$Al$_7$Si$_5$}

\author{R. Guehne}
\email{robin.guehne@cpfs.mpg.de}
\author{A. K. Sharma}
\author{P. Yanda}
\email{Premakumar.Yanda@lmu.de}
\author{J. Noky}
\author{J. Sichelschmidt}
\author{R. Koban}
\author{W. Schnelle}
\author{C. Shekhar}
\author{M. Baenitz}
\author{C. Felser}
\affiliation{Max Planck Institute for Chemical Physics of Solids, 01187 Dresden, Germany}

\begin{abstract}
The metallic kagome compound Sc$_3$Mn$_3$Al$_7$Si$_5$ has attracted attention as a candidate platform where geometric frustration and itinerant electrons may cooperate to stabilize a quantum-disordered magnetic ground state. Here, we combine bulk thermodynamic probes, low-noise FIB-device transport, and comprehensive $^{55}$Mn Nuclear Magnetic Resonance (NMR) measurements to elucidate the low-temperature spin dynamics of this system. The bulk data reveal strongly reduced magnetic entropy, a negative magnetoresistance arising from spin scattering, and field-dependent transport indicates the spin fluctuations, while showing no signatures of long-range magnetic order. NMR provides a direct local view of the correlated Mn moments: the nuclear spin-spin relaxation $T_2$ exhibits a pronounced low-temperature enhancement driven by an indirect internuclear coupling through electronic spin fluctuations, whose temperature and distance dependence point to partially gapped low-energy spin excitations. The spin-lattice relaxation rate $T_1^{-1}$ displays a Hebel-Slichter-like coherence peak near \SI{10}{K}, coincident with the resistivity crossover and a subtle heat-capacity anomaly, indicating the formation of short-range spin-singlet correlations. Together, our results demonstrate that Sc$_3$Mn$_3$Al$_7$Si$_5$ hosts an unconventional correlated state dominated by frustrated, gapped spin dynamics, placing it among the rare metallic kagome systems proximate to a quantum spin liquid.
\end{abstract}

\maketitle

\section{Introduction}
Quantum spin liquids (QSLs) are exotic states of matter in which geometric frustration and quantum fluctuations prevent spins from developing long-range magnetic order, even at the lowest temperatures \cite{Lacroix2011,Meng2012,Savary2016,Zhou2017,Wen2019,Broholm2020}. The kagome lattice, a planar network of corner-sharing triangles, provides an ideal stage for realizing such states, as vividly demonstrated in insulating compounds with Cu$^{2+}$ (spin-$1/2$) layers, where fractionalized excitations and continuum spin dynamics have been established \cite{Helton2007,Okamoto2007,Mendels2016,Fu2015,Norman2016,Liao2017}. In metallic systems with kagome layers, the situation is even more intriguing: flat bands, Dirac dispersions, and strong spin correlations intertwine with itinerant electrons, enabling unconventional charge transport coupled to frustrated magnetism \cite{Sharma2025}. For kagome metals, such as FeSn and AV$_3$Sb$_5$, it has been elucidated how lattice geometry can stabilize novel correlated states, from topological magnetism to charge order \cite{Ortiz2020,Kang2020}. The recently reported compound HoAgGe, a distorted kagome material, is claimed to host a spin-ice state arising from frustration of Ising-like Ho moments on the kagome network \cite{Zhao2020}. 
A central challenge, however, remains to identify disorder-free metallic kagome systems in which geometric frustration can stabilize short-range spin-singlet correlations, the metallic analogue of the singlet physics often associated with insulating quantum spin liquids \cite{Wang2021}. Such singlet formation in a metal is rare, yet theoretically anticipated, motivating the search for kagome metals where frustration suppresses classical magnetism and allows spin-singlet or gapped-fluctuation states to develop.

The kagome metal Sc$_3$Mn$_3$Al$_7$Si$_5$ has recently discussed as candidate for Mn-based QSL \cite{He2014}. Initial work established its synthesis and crystal structure, together with bulk magnetic and transport properties that hinted at correlated behavior \cite{Samanta2024}. Magnetic, heat capacity, and neutron scattering measurements established the absence of long range magnetic order and revealed a markedly reduce Mn moment, consistent with an itinerant magnetic character on the kagome network, whereas subsequent theoretical and spectroscopic work suggested orbital-selective flat bands and concomitant ferromagnetic fluctuations \cite{He2014,Li2021,Samanta2024,Ding2025}. These features naturally raise the question of whether short-range spin singlets or a gapped spin-fluctuation regime may form at low temperature in this metal.

Here, we present a comprehensive investigation of high-quality single crystals, combining thermodynamic measurements with electrical transport and Hall effect on a device cut by focused ion beam (FIB), as well as electron spin resonance (ESR) and a detailed $^{55}$Mn nuclear magnetic resonance (NMR) study. Our results demonstrate suppressed magnetic entropy, a field dependent resistivity crossover coincident with a heat capacity anomaly, negative magnetoresistance driven by spin scattering, and field-dependent transport signatures of strong spin fluctuations. While ESR remains silent, nuclear spins are indirectly coupled via electronic spin fluctuations yielding a unique nuclear spin-spin relaxation below about \SI{50}{K} beside evidence of a Hebel-Slichter coherence peak implying a temperature dependent spin-gap and spin-singlet formation. These findings establish Sc$_3$Mn$_3$Al$_7$Si$_5$ as a compelling metallic kagome system in which geometric frustration and itinerancy cooperate to generate short range singlet physics.

\section{Results and Discussion}

\begin{figure}[t]
\centering
\includegraphics[width=\textwidth]{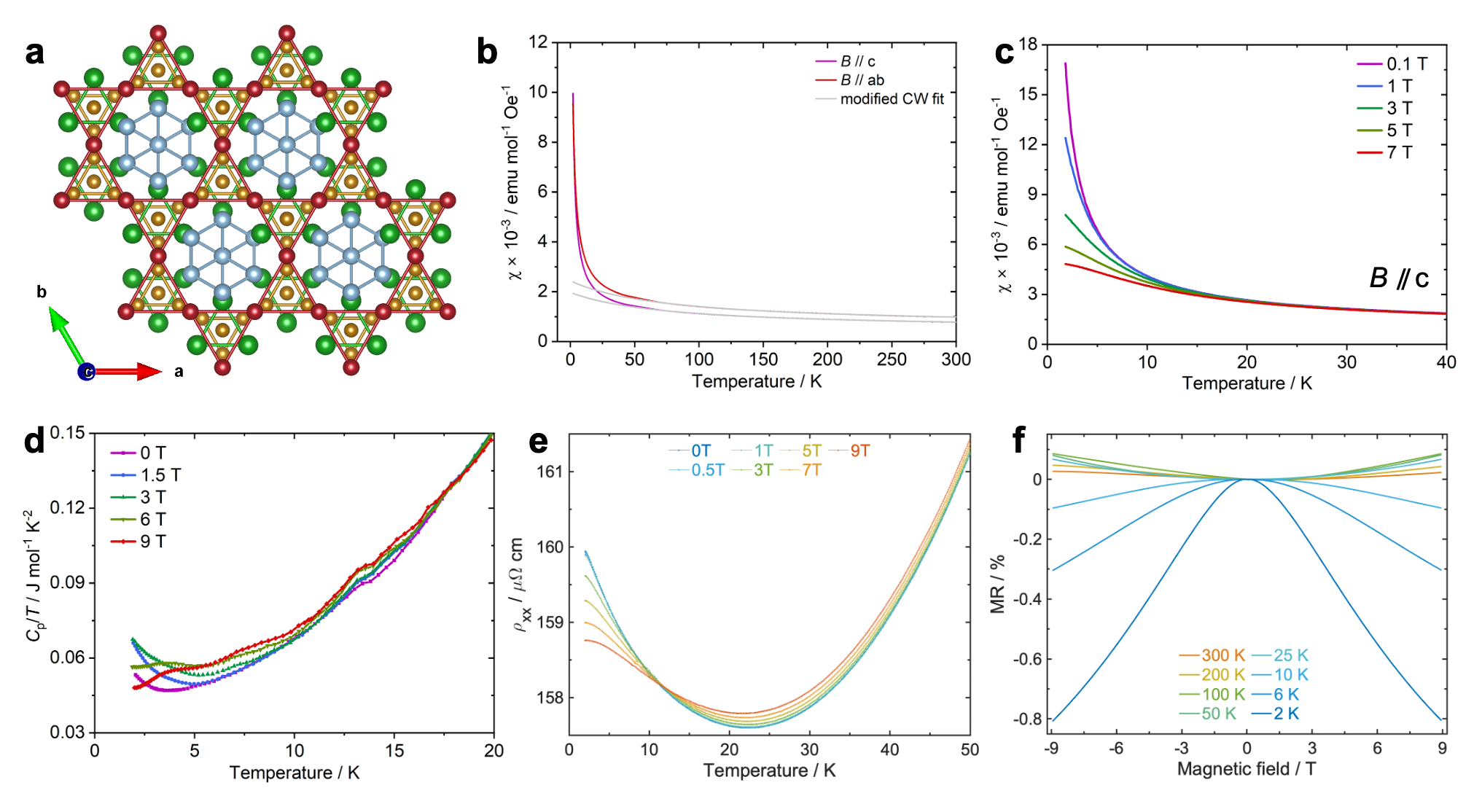}
\caption{\footnotesize{\label{Fig0}
(\textbf{\textsf{a}}) Crystal structure of Sc$_3$Mn$_3$Al$_7$Si$_5$,
(\textbf{\textsf{b}}) Temperature dependent DC susceptibility $\chi(T)$ and modified CW with the field along $c$ and $ab$.
(\textbf{\textsf{c}}) Temperature dependent DC susceptibility $\chi(T)$ at different fields for $B_0\parallel c$.
(\textbf{\textsf{d}}) $C_p/T$ versus $T$ in the presence of various fields.
(\textbf{\textsf{e}}) Details of the temperature dependent longitudinal resistivity $(\rho_{\mathrm{xx}}$) at various fields. See Fig.~\ref{FigS04} in the SI for the full temperature range.
(\textbf{\textsf{f}}) Field dependence of magnetoresistance $\textrm{MR}(\%)=100\times(\rho-\rho(0)) / \rho(0)$  at various temperatures for $I\parallel ab$ and $B\parallel c$.
}}
\end{figure}

Sc$_3$Mn$_3$Al$_7$Si$_5$ crystallizes in a hexagonal lattice with $P$6$_3/mmc$ ($No.\ 194$) space group \cite{He2014}. The crystal structure consists of Mn atoms forming a planar kagome network separated by Sc and Al/Si layers (Fig.~\ref{Fig0}\textbf{a}), providing a geometrically frustrated framework for magnetic interactions.

The DC magnetic susceptibility $\chi(T)$ reveals no signature of long-range magnetic order down to 2 K (Fig.~\ref{Fig0}\textbf{c}). Instead, $\chi(T)$ increases strongly upon cooling, with the divergence most pronounced at low fields and progressively suppressed by the applied field. A modified Curie-Weiss analysis to $\chi(T)$, as shown in Fig.~\ref{Fig0}\textbf{b}, yields $\chi_0=\SI{4.8E-4}{emu\ mol^{-1}Oe^{-1}}$, an effective Mn moment of $\mu_{\mathrm{eff}}=\SI{1.0(1)}{\mu_{\mathrm{B}}/Mn}$, a Curie-Weiss temperature $\theta_\mathrm{CW}=\SI{-76(2)}{K}$ for $B\parallel c$ and $\chi_0=\SI{6.1E-4}{emu\ mol^{-1}Oe^{-1}}$, $\SI{1.0(1)}{\mu_{\mathrm{B}}/Mn}$, and $\theta_\mathrm{CW}=\SI{-77(2)}{K}$ for $B\parallel ab$. 
The temperature-independent susceptibility $\chi_0$ is sizable for both orientations, comparable to previous reports \cite{He2014,Samanta2024}. This likely reflects enhanced Pauli- or Van Vleck-contributions arising from the flat-band–enhanced density of states. Other factors like tiny amounts of localized impurities and weak ferromagnetic moments cannot entirely be ruled out within the present study.
The Curie-Weiss response is essentially isotropic, in agreement with earlier work. The small effective Mn moment indicates that the Mn 3$d$ electrons are strongly hybridized and the magnetism is predominantly itinerant in nature. 
The negative $\theta_\mathrm{CW}$ indicates dominant antiferromagnetic interactions or correlated magnetism \cite{Samanta2024,Ding2025}. The absence of long-range magnetic ordering despite large $\theta_\mathrm{CW}$ points to strongly fluctuating character of spin interactions on the Mn kagome lattice. After subtracting the linear $M(B)$ background at \SI{20}{K}, the isothermal magnetization at \SI{2}{K} (Fig.~\ref{FigS02}) reveals a weak nonlinear contribution that saturates by $\sim\SI{5}{T}$ for both $B\parallel c$ and $B\parallel ab$. The low value of the saturated magnetization indicates the presence of fluctuating spins. The nearly isotropic response and absence of hysteresis indicate the polarization of fluctuating short-range correlated Mn moments rather than the onset of long-range order.

The specific heat $C_{\mathrm{p}}(T)$ of Sc$_3$Mn$_3$Al$_7$Si$_5$ shows no sharp anomaly down to 2 K, ruling out the presence of conventional long-range magnetic order (Fig.~\ref{FigS03}\textbf{a}). To estimate the magnetic contribution $C_{\mathrm{mag}}(T)$, the lattice contribution modeled using a combined Debye–Einstein expression, which is a standard approach for complex intermetallics in the absence of a nonmagnetic analogue. Although such a phonon model is necessarily approximate, it provides a reasonable baseline from which to assess the overall scale of magnetic contributions. The resulting magnetic entropy per formula unit, obtained by integrating $C_{\mathrm{mag}}(T)$, remains well below \SI{1}{J\ mol^{-1}K^{-1}} up to \SI{50}{K}, which is less than \SI{1}{\%} of the full theoretical value $R\ln(2S+1)$ expected for localized Mn$^{3+}$ moments $(S = 2)$. Regardless of the exact phonon parameterization, only a very small fraction of the expected magnetic entropy is released at low temperatures. Such strong entropy suppression is characteristic of systems where magnetic degrees of freedom are either strongly fluctuating and itinerant.
At the lowest temperatures, $C_{\mathrm{p}}/T$ exhibits an upturn that which changes characterstically with field and is captured by a Schottky term, consistent with nuclear contributions rather than intrinsic order. Surprisingly, a subtle shoulder appears near $T^{\ast}\approx\SI{12}{K}$, as shown in Fig.~\ref{Fig0}\textbf{d}, which is not observed previously \cite{Samanta2024}. This feature, coincident with anomalies in both resistivity and NMR relaxation, which will be discussed later, likely reflects a crossover in the spin excitation spectrum. At low temperatures, the $C_p/T$ versus $T^2$ data exhibits a linear regime between the Schottky anomaly and the characteristic temperature $T^{\ast}$. A fit in this range (Fig.~\ref{FigS03}\textbf{c}) gives a linear coefficient $\gamma_\mathrm{e} =\SI{40}{mJ\ mol^{-1}K^{-2}}$, indicative of substantial low-energy electronic correlations, and a phonon coefficient $\beta=\SI{0.276}{mJ\ mol^{-1}K^{-4}}$, corresponding to a Debye temperature $\Theta_\mathrm{D}\sim\SI{500}{K}$. The coexistence of a finite $\gamma_\mathrm{e}$ with suppressed entropy suggests a correlated metallic state, in which Mn moments remain fluctuating rather than ordering.

Electrical transport was measured exclusively on micro structured devices prepared using focused ion beam (FIB) technique (Fig.~\ref{FigS04}\textbf{a}). The resistivity $\rho(T)$ decreases monotonically upon cooling, consistent with poor metallic behavior at high temperatures. A large residual resistivity $\rho_0$, consistent with earlier reports \cite{He2014,Samanta2024}, likely reflects intrinsic alloy scattering in this multicomponent metallic system. Interestingly, $\rho(T)$ develops an upturn below $\sim\SI{40}{K}$ (Fig.~\ref{FigS04}\textbf{b}). This low temperature upturn may be due to enhanced scattering from spin fluctuations, which act to impede charge transport in the absence of long-range magnetic order. Application of magnetic fields systematically suppresses the upturn and reveals a characteristic crossover near $T^{\ast} \sim\SI{12}{K}$, coincident with the anomaly observed in heat capacity (Fig.~\ref{Fig0}\textbf{e}). The negative magnetoresistance (MR) seen at 2, 6, 10, and \SI{15}{K} points to the suppression of spin-disorder scattering by field-induced spin polarization (Fig.~\ref{Fig0}\textbf{f}). At higher temperatures, the MR changes sign and becomes positive, reflecting the dominance of conventional orbital contributions once magnetic fluctuations are less effective scatterers. Hall effect measurements further support this picture (Fig.~\ref{FigS04}\textbf{c}). While both the carrier density and mobility remain nearly constant below \SI{40}{K}, they exhibit pronounced variations above this temperature, consistent with a change in the dominant scattering mechanism (Fig.~\ref{FigS04}\textbf{d}). The combined evolution of resistivity, magnetoresistance, and Hall response underscores the role of fluctuating Mn moments in controlling transport and highlights the strong interplay between spin correlations and conduction electrons in Sc$_3$Mn$_3$Al$_7$Si$_5$.

To directly probe these spin dynamics and test their microscopic origin, we employ $^{55}$NMR NMR and ESR (electron spin resonance) as local spin probes. ESR measurements were performed at \SI{9.4}{GHz} (X-band) corresponding to a resonance field of the Mn spins at around \SI{0.3}{T}, assuming $g_{\mathrm{Mn}}=2$. Interestingly, we found no resonance absorption in the whole temperature range investigated (\si{3}-\SI{300}{K}). We suspect that this ESR silence is due to a large resonance-linewidth originating from strong Mn-spin fluctuations which are present in the whole T-range as indicated by susceptibility and specific heat. The effect of spin fluctuations on the ESR linewidth has been exemplified, for instance, for the weak itinerant ferromagnet MnSi on the basis of Moriya’s theory for spin fluctuations in itinerant electron magnets \cite{Demishev2011,Moriya2012}.

\begin{figure}[t]
\centering
\includegraphics[width=.5\textwidth]{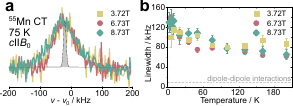}
\caption{\footnotesize{\label{Fig1}
(\textbf{\textsf{a}}) The $^{55}$Mn central transition (CT) in frequency units (kHz) at \SI{75}{K} for 3 different magnetic fields applied parallel to the $c$-axis. The gray line shows the approximate line broadening as expected from direct dipole-dipole interactions ($\sim\SI{10}{kHz}$) \cite{Vleck1948}.
(\textbf{\textsf{b}}) Temperature dependent linewidth (full width at half maximum) in frequency units of the $^{55}$Mn CT for 3 different fields.
}}
\end{figure}

NMR of Sc$_3$Mn$_3$Al$_7$Si$_5$ has so far only been reported for $^{27}$Al nuclei, while there is no account of $^{55}$Mn NMR, even though such measurements offer a unique microscopic probe of the intrinsic spin fluctuations responsible for the correlated behavior in this kagome metal. \cite{Samanta2024,Ding2025}.
Representative spectra of the $^{55}$Mn central transition (CT) related to transitions between the nuclear $\pm1/2$ spin states are shown in Fig.~\ref{Fig1}\textbf{a}. The spectra are provided in frequency units (kHz), the respective carrier frequencies of the measurements have been subtracted (the moderate and negative NMR shift is discussed in the supplementary information (SI)). The significance of the spectra in Fig.~\ref{Fig1}\textbf{a} lies in their large and field independent linewidths (80\SI{-90}{kHz}). 
Field independent line broadening cannot be assigned to defects and corresponding distributions of the magnetic ($\propto B_0$) and the $2^{\mathrm{nd}}$ order electric quadrupole shift ($\propto B_0^{-1}$).  
The linewidth increases with decreasing temperatures (panel \textbf{b}) and remains essentially independent of the external field over the full temperature range. 
Field independent broadening typically stems from direct dipole-dipole interactions, which, however, are much too small to account for the observed broadening. Therefore, nuclei must be coupled indirectly via electronic excitations \cite{Ruderman1954,Bloembergen1955}.
Through such an indirect internuclear coupling, contributions from neighboring nuclear magnetic moments to the local magnetic field seen by a nucleus are significantly amplified, yielding the observed broadening.

\begin{figure}[t]
\centering
\includegraphics[width=.9\textwidth]{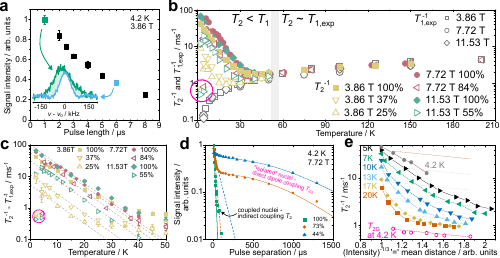}
\caption{\footnotesize{\label{Fig2}
(\textbf{\textsf{a}}) Signal intensity ($T_2$ corrected) for various pulse lengths (excitation band widths) and two example resonances for short ($\SI{1}{\mu s}$) and long ($\SI{6}{\mu s}$) pulses.
(\textbf{\textsf{b}}) The spin-spin relaxation rate ($T_2^{-1}$) and inverse CT lifetime ($T_{1,\mathrm{exp}}^{-1}$) as functions of temperature for the 3 external fields \SI{3.86}{T}, \SI{7.72}{T}, and \SI{11.53}{T}. The grey bar separates the two temperature regimes (see text for details). The very small relaxation rates at the lowest temperatures (pink circle) denote isolated spins that decay via direct dipolar couplings.
(\textbf{\textsf{c}}) Spin-spin relaxation rate after subtraction of the corresponding CT lifetime. Grey parallel lines are guides to the eye, corresponding to an exponential function with $\tilde{T}=\SI{7.1}{K}$.
(\textbf{\textsf{d}}) $T_2$ measurement (signal intensity as function of pulse delay) for \SI{4.2}{K}, \SI{7.73}{T}, and various excitation conditions (\SI{100}{\%}, \SI{73}{\%}, and \SI{44}{\%} excited nuclei). Solid lines are fittings using a sum of a single exponential ($T_2$) and a Gaussian decay ($T_{\mathrm{2G}}$).
(\textbf{\textsf{e}}) The spin-spin relaxation rates as a function of the mean distance between Mn nuclei for various temperatures.
}}
\end{figure}

Consequently, the Mn resonance is governed by a substantial inhomogeneous broadening due to the fields of near neighbors, including $^{45}$Sc (Natural abundance $NA=\SI{100}{\%}$), $^{27}$Al ($NA=\SI{100}{\%}$), and other Mn nuclei (less so $^{29}$Si due to the low $NA<\SI{5}{\%}$). This allows us to limit the number of excited nuclei during a measurement. The pulse in an NMR experiment acts as a band pass with a narrow frequency window inversely proportional to the pulse length \footnote{The Fourier transform of the rectangular pulse of length $\tau$ is a sinc function, $\sin{(\omega\tau})/\omega$, with a full width at half maximum of $1/\tau$}. As shown in Fig.~\ref{Fig2}\textbf{a}, by increasing the exciting pulses' lengths, we systematically narrow down the observed resonance line and thus, the corresponding signal intensity (area under the curve) which is proportional to the number of excited nuclei. As will been shown next, this severely affects the low temperature spin-spin relaxation ($T_2$) of Mn nuclei which is the timescale of phase coherence of the nuclear precession.

\begin{figure}[t]
\centering
\includegraphics[width=.5\textwidth]{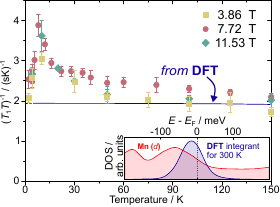}
\caption{\footnotesize{\label{Fig3}
 (\textbf{\textsf{a}}) The spin-lattice relaxation rate divided by the temperature $(T_1T)^{-1}$ as a function of temperature for 3 different magnetic fields, including the calculated behavior from the DOS (blue).
(\textit{inset}) The density of states (DOS) for Mn $d$-orbitals (red) near the Fermi level. To extract the temperature dependence of the spin-lattice relaxation, we calculated $T_1^{-1}\propto\int f(E-\mu_c)[1-f(E-\mu_c)]D(E)^2\mathrm{d}E$, using the Fermi function $f(E-\mu_c)$ for various temperatures. The blue line denotes the integrant for \SI{300}{K} representing the range of thermally excited states available for nuclear relaxation. 
 }}
\end{figure}

In Fig.~\ref{Fig2}\textbf{b}, we present ($T_2^{-1}$) of $^{55}$Mn nuclei for a variety of external fields and excitation conditions. 
Also shown in the figure and denoted by the gray open symbols is the temperature dependent inverse CT lifetime $T_{\mathrm{1,exp}}^{-1}$, i.e., the effective recovery rate which is directly related to the system's spin-lattice relaxation rate $T_1^{-1}$ \footnote{Note, the apparent lifetime of individually excited transitions is much faster than the lifetime of the entire system which was presented in Fig.~\ref{Fig3}. The CT lifetime is obtained from approximating the apparent recovery of the CT with a single exponential (effective two-level system).}. 
We identify two regimes. For temperatures above about \SI{50}{K}, $T_2^{-1}$ equals approximately $T_{\mathrm{1,exp}}^{-1}$. This implies that the very short lifetime sets a natural limit for the phase coherence of the nuclear precession in which case one expects that $T_2=2T_1$, or, as related to the ESR silence, fast and isotropic spin fluctuations govern both, spin-spin as well as spin-lattice relaxation. For $T<\SI{50}{K}$, however, the $T_2^{-1}$ values deviate strongly from $T_{\mathrm{1,exp}}^{-1}$, implying the spin-spin relaxation to be driven by a different mechanism. 
An effective spin-echo decay ($T_2$) is observed if coupled nuclei are flipped simultaneously during a measurement, because this will change their mutual local field causing a loss of coherence. 
Therefore, a strong indirect coupling as reflected in the extensive line broadening will cause fast spin-spin relaxation (as in cuprates \cite{Pennington1991}). 
The temperature dependence implies that the indirect coupling requires virtual transition across a very small gap and is thus of a Bloembergen-Rowland (BR) rather than a Ruderman-Kittel-Kasuya-Yoshida-type (RKKY) which involves transitions between degenerate bands \cite{BR1955,Ruderman1954}. Thermal excitations interfere with the virtual transitions and lead to the exponential suppression of $T_2^{-1}$ with temperature as can be seen in Fig.~\ref{Fig3}\textbf{c}. 
We find $\tilde{T}\approx \SI{7.1}{K}$ as the characteristic temperature from a simple exponential fit, corresponding to \SI{0.61}{meV} (\SI{148}{GHz}) for the spin excitations.

We repeated the $T_2$ measurements under various excitation conditions.
While the solid symbols in Fig.~\ref{Fig3}\textbf{b} and \textbf{c} represent measurements where the full resonance line was excited with sufficiently short pulses (broad excitation width), i.e., \SI{100}{\%} of the CT nuclei, the open triangles denote measurements with longer pulses/reduced bandwidths and thus fewer excited nuclei. 
The effect on $T_2^{-1}$ is substantial. While the external field seems not to have any significant impact on the low temperature $T_2^{-1}$ behavior (solid symbols in panel \textbf{a} and \textbf{b}), reducing the number of excited nuclei shifts the upturn of $T_2^{-1}$ systematically to lower temperatures (open triangles), from initially $\sim$\SI{50}{K} for \SI{100}{\%} of the nuclei to less than \SI{25}{K} for about \SI{25}{\%} (yellow triangles). 
Reducing the number of excited nuclei during an experiment for a given temperature (triangles in Fig.~\ref{Fig2}\textbf{b} and \textbf{c}) means to look at an increasingly dilute spin system with the mean distance between excited spins increasing, because any fraction of excited nuclei will fill the same volume. 
And since the underlaying electronic correlation decays rapidly in real-space, the corresponding internuclear coupling is increasingly hampered, yielding $T_2^{-1}$ values to fall as the spin system is further diluted. 
In other words, by reducing the number of excited nuclei, we gradually decouple them from each other which is reflected in a longer spin-spin relaxation. 
That the spin-spin relaxation can be manipulated in this way shows that slow ferromagnetic fluctuations as a direct source can be ruled out as these would yield transverse relaxation \emph{independent} of the way the spin-system is perturbed.

At the lowest temperatures (2.1 and \SI{4.2}{K}), up to $25-\SI{30}{\%}$ of the Mn nuclei can even be entirely decoupled, yielding the very small decay rates highlighted by pink circles in Fig.~\ref{Fig2}\textbf{b} and \textbf{c} (see also SI). This slow decay (Gaussian $T_{\textrm{2G}}$ in panel \textbf{d}) stems from direct dipole-dipole interactions that are weak but longer ranging ($\sim r^{-3}$) than the indirect coupling. This observation proves the indirect coupling to decay very rapidly in real space.
To track the distance dependence of the indirect coupling, we conducted systematic temperature and excitation dependent measurements and plot $T_2^{-1}$ vs. the inverse cube root signal intensity, cf. Fig.~\ref{Fig2}\textbf{e}. The latter is related to the mean distance between coupling partners for an isotropic system in 3 dimensions. 
At higher temperatures, the decay rapidly levels off due to the lifetime of the nucleus and its (unlike) neighbors offering efficient relaxation channels.
At low temperatures these contributions freeze out, leaving the distance dependence to reflect only the indirect coupling. 
The decay is much steeper than that of the dipolar component at \SI{4.2}{K} (pink data points and gray dashed lines) and together with the occurrence of decoupled nuclei confirms a very short range, perhaps an exponential distance dependence as expected for gapped spin fluctuations.

Finally, we address the spin-lattice relaxation $T_1$ of $^{55}$Mn nuclei. In metallic system, one typically analyzes the density of states (DOS) to derive the temperature dependence of the corresponding spin-lattice relaxation. A peak in the Mn DOS as expected from a flat band near the Fermi level gives rise to a slight downturn in the corresponding $(T_1T)^{-1}$ for high temperatures, but is too far from $E_0$ ($\sim\SI{-80}{meV}$) to affect the essential low temperature behavior of the spin-lattice relaxation, as shown in Fig.~\ref{Fig3} by the blue line.
Instead, we observe a clear peak in $(T_1T)^{-1}$ near \SI{10}{K}, typically suggesting a magnetic phase transition, as in the case of the itinerant ferromagnet ZrZn$_2$ \cite{Kontani1977}, but it lacks an inverse field dependence. With the absence of long-range order from bulk measurements and in the light of strongly coupled nuclei via potentially gapped spin fluctuations as reflected in the spin-spin relaxation, this peak could be read as a Hebel-Slichter coherence peak, implying a low energy spin-gap and a singlet ground state \cite{Hebel1959}. The weakness and width of the peak is in accordance with low temperature spin-spin relaxation behavior suggesting a continuous singlet formation spread over a large temperature range rather than a well defined phase transition. Notably, the $T_1^{-1}$ peak coincides with the subtle anomaly in heat capacity at $T^{\ast}\sim\SI{12}{K}$ and with the resistivity crossover, reinforcing a unified picture of evolving singlet-like correlations at low temperature.

\section{Conclusions}
\noindent
In summary, we present a comprehensive investigation of the kagome metal Sc$_3$Mn$_3$Al$_7$Si$_5$ through bulk thermodynamic and transport measurements together with local spin probes.
The NMR spin–spin relaxation $T_2$ reveals a pronounced low-temperature enhancement arising from an indirect internuclear coupling via electronic spin fluctuations, whose temperature and spatial dependence indicate that gapped spin excitations become relevant below $\sim\SI{50}{K}$.
A corresponding Hebel-Slichter peak in the spin-lattice relaxation around \SI{10}{K}, coincident with subtle anomalies in heat capacity and resistivity, points to the formation of short range spin-singlet correlations in the absence of long range magnetic order.
Rather than undergoing a conventional phase transition, Sc$_3$Mn$_3$Al$_7$Si$_5$ thus enters a quantum-disordered regime dominated by frustrated, slowly fluctuating spin singlets.
Our results highlight the need for a theoretical framework for the indirect nuclear coupling in the presence of strong spin fluctuations and demonstrate that $T_2$ relaxation may provide a sensitive tool to uncover hidden magnetic dynamics in correlated quantum materials.

\section*{Methods}

\noindent
\textbf{Crystal synthesis and characterization}  Single crystals of Sc$_3$Mn$_3$Al$_7$Si$_5$ (SMAS) were successfully grown using the Al-self flux method with a ratio of Sc:Mn:Al:Si $\sim$ 3:3:30:5. All the elements were of high purity, and thus, big single crystals were grown. They were found to be stable in the air and moisture. The magnetization measurements were performed using a Quantum Design MPMS-3. Resistivity, magnetoresistance, and Hall measurements were carried out on a device made by Thermo Scientific Helios 5 CX. The device contained 8 contacts, and all were used at the same time. The size of the Hall bar was $76\times 9.8 \times 4.7$mm$^3$. The heat capacity was measured using the HC option of a PPMS.\\

\noindent
\textbf{NMR} Measurements were carried out on \textsc{Janis} sweepable \SI{9}{T} and \SI{14}{T} magnets using a \textsc{Tecmag Apollo} NMR console. Two flat ($\sim1\times1\times\SI{0.3}{mm^3}$) and hexagonal samples were cut from a large single crystal and placed in a high (S1 - 8 turn) and a low (S2 - 20 turns) frequency coil. All experiments were carried out with $c\parallel B_0$. Echo experiments ($\pi/2-\tau-\pi$) were used to excite individual transitions, their $\tau$-dependence was studied for transverse ($T_2$) and their recovery following a $\pi/2$-saturation pulse for longitudinal relaxation ($T_1$). The shift was obtained by referencing the respective spin species via the rf-coil's $^{63}$Cu resonance line and the second reference method \cite{Harris2008}.\\

\noindent
\textbf{ESR} Electron spin resonance studies on platelet and needle shaped crystals (1.25mg and 11.5mg) were performed at X-band frequency, 9.4 GHz, using a standard continuous-wave spectrometer for temperatures between 3.5K and 300K. \\

\noindent
\textbf{Numerical calculations} The simulated results were obtained by using \textit{ab initio} calculations in the framework of density-functional theory (DFT), as implemented in the program VASP~\cite{kresse1996}. In this code, augmented plane waves are used as a basis set together with pseudopotentials. To describe the exchange-correlation potential, the generalized-gradient approximation (GGA)~\cite{perdew1996} was used.

The self-consistent calculations were carried out on a $20 \times 20\times 10$ k mesh. Convergence for total energy and DOS was carefully checked. For the DOS calculations, a k mesh of $60\times 60\times 30$ was used.

\section*{Acknowledgment}

\noindent
R.G. is particularly grateful for discussions with B. Fine and J. Haase (Leipzig). The authors thank O. Stockert, S. Wirth, and M. Brando. We acknowledge the financial support by the Deutsche Forschungsgemeinschaft (DFG) under SFB1143 (Project No. 247310070), the Würzburg-Dresden Cluster of Excellence on Complexity and Topology in Quantum Matter ct.qmat (EXC 2147, Project No. 390858490).

\section*{Author contributions}
\noindent
P.Y. grew the single crystals. P.Y. and A.K.S. performed all bulk thermodynamic measurements (magnetization and heat capacity), and carried out the magnetotransport experiments with assistance from R.K., W.S., and C.S., and analyzed the corresponding data. A.K.S did the FIB device fabrication for transport measurements. R.G. conducted and analyzed NMR experiments with assistance from M.B., J. Sichelschmidt carried out ESR experiments, J.N. carried out DFT calculations, C.F. lead the project, P.Y. planned, executed, and wrote the bulk properties section of the manuscript with the help of A.K.S. R.G. prepared the overall manuscript with contributions from all authors and substantial input from P.Y.

\bibliography{references}

\newpage

\appendix

\setcounter{figure}{0}
\renewcommand{\thefigure}{S\arabic{figure}}

{\Huge{Supplementary Information}}\\

\section*{Single crystal samples}
Single crystals of Sc$_3$Mn$_3$Al$_7$Si$_5$ were grown using the flux method, yielding millimeter-sized, hexagonal, rod-like crystals suitable for transport, magnetization, and local-probe measurements. 

\begin{figure}[h]
\centering
\includegraphics[width=.5\textwidth]{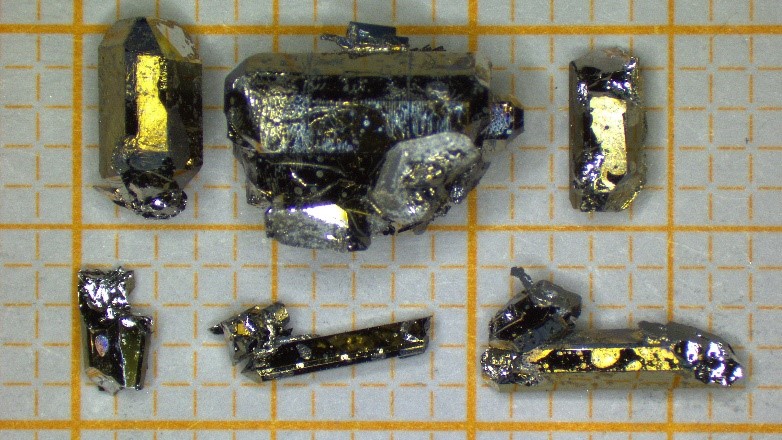}
\caption{\footnotesize{\label{FigS01}
Photographic image of as grown crystals of Sc$_3$Mn$_3$Al$_7$Si$_5$ single crystals.
}}
\end{figure}

\section*{Magnetization, Heat capacity, AC susceptibility, and Transport measurements}

\begin{figure}[h]
\centering
\includegraphics[width=.4\textwidth]{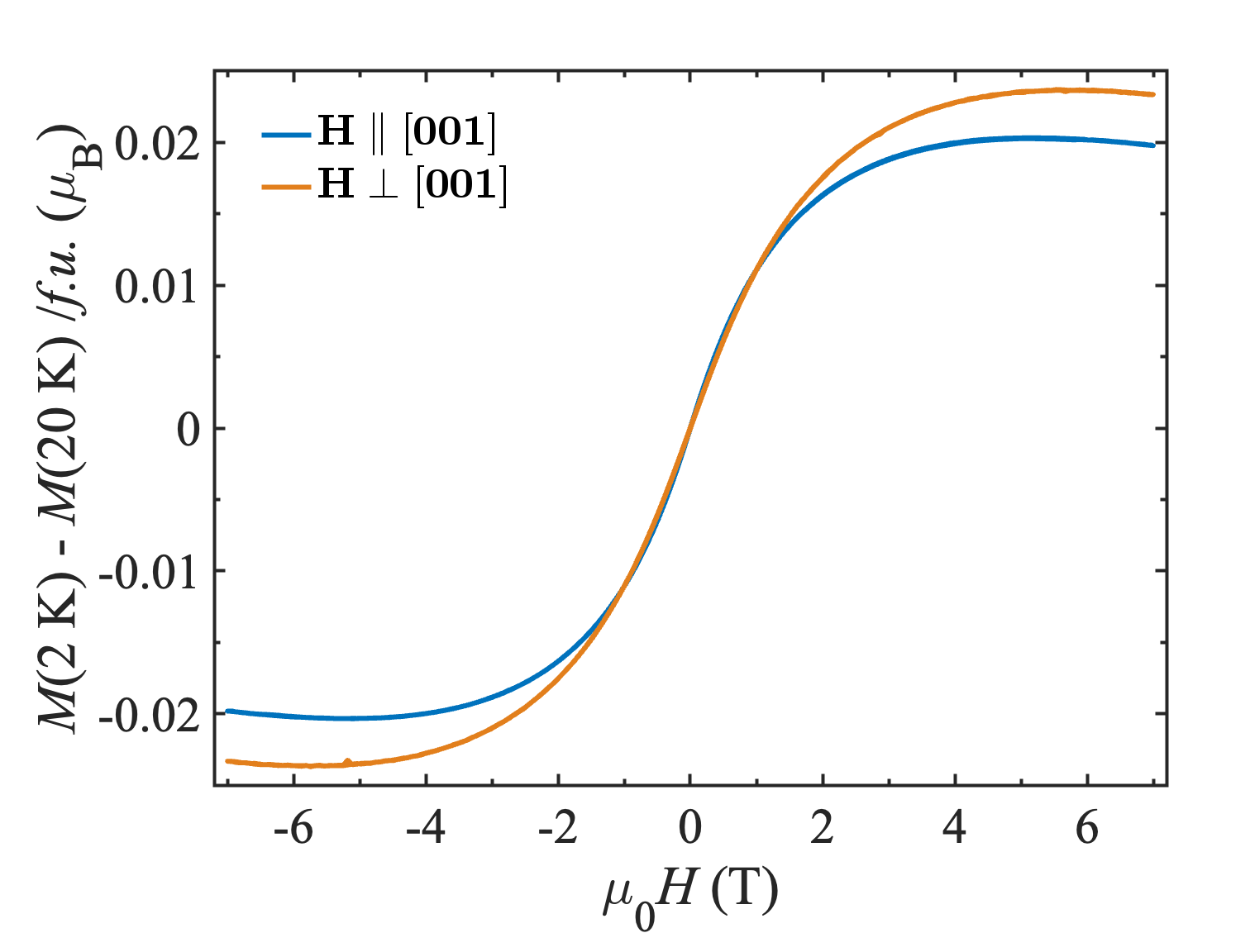}
\caption{\footnotesize{\label{FigS02}
Difference of the isothermal magnetization curves taken at $T = \SI{2}{K}$ and $T = \SI{20}{K}$ for two magnetic field directions with respect to the crystallographic $c$ axis.
}}
\end{figure}

\begin{figure}[h]
\centering
\includegraphics[width=\textwidth]{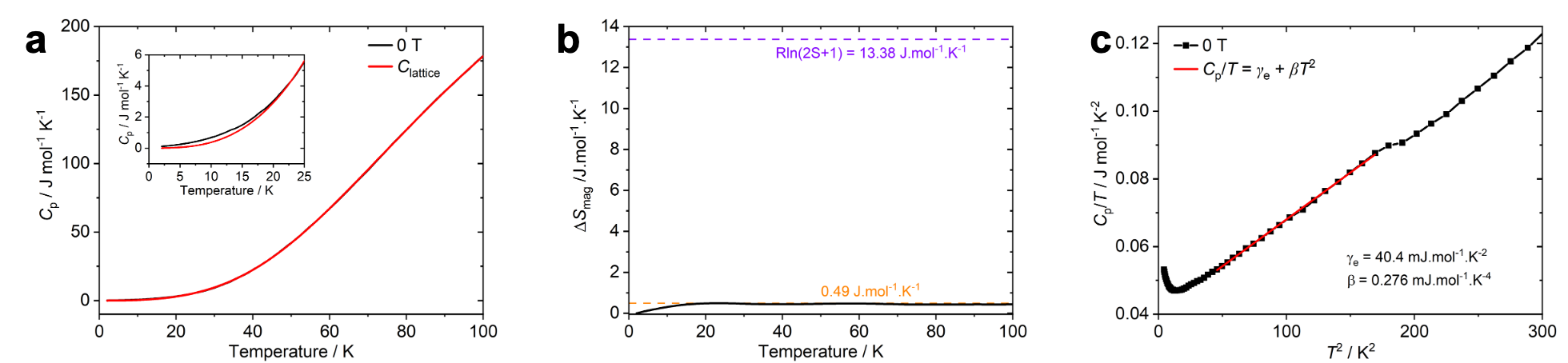}
\caption{\footnotesize{\label{FigS03}
(\textbf{\textsf{a}}) Heat Capacity (Cp) versus temperature at 0 T and its lattice contribution,
(\textbf{\textsf{b}}) Magnetic entropy extracted by subtracting the lattice contribution, 
(\textbf{\textsf{c}}) $C_{\mathrm{p}}p/T$ versus $T^2$ at low temperatures and its fit.
}}
\end{figure}

\begin{figure}[h!]
\centering
\includegraphics[width=\textwidth]{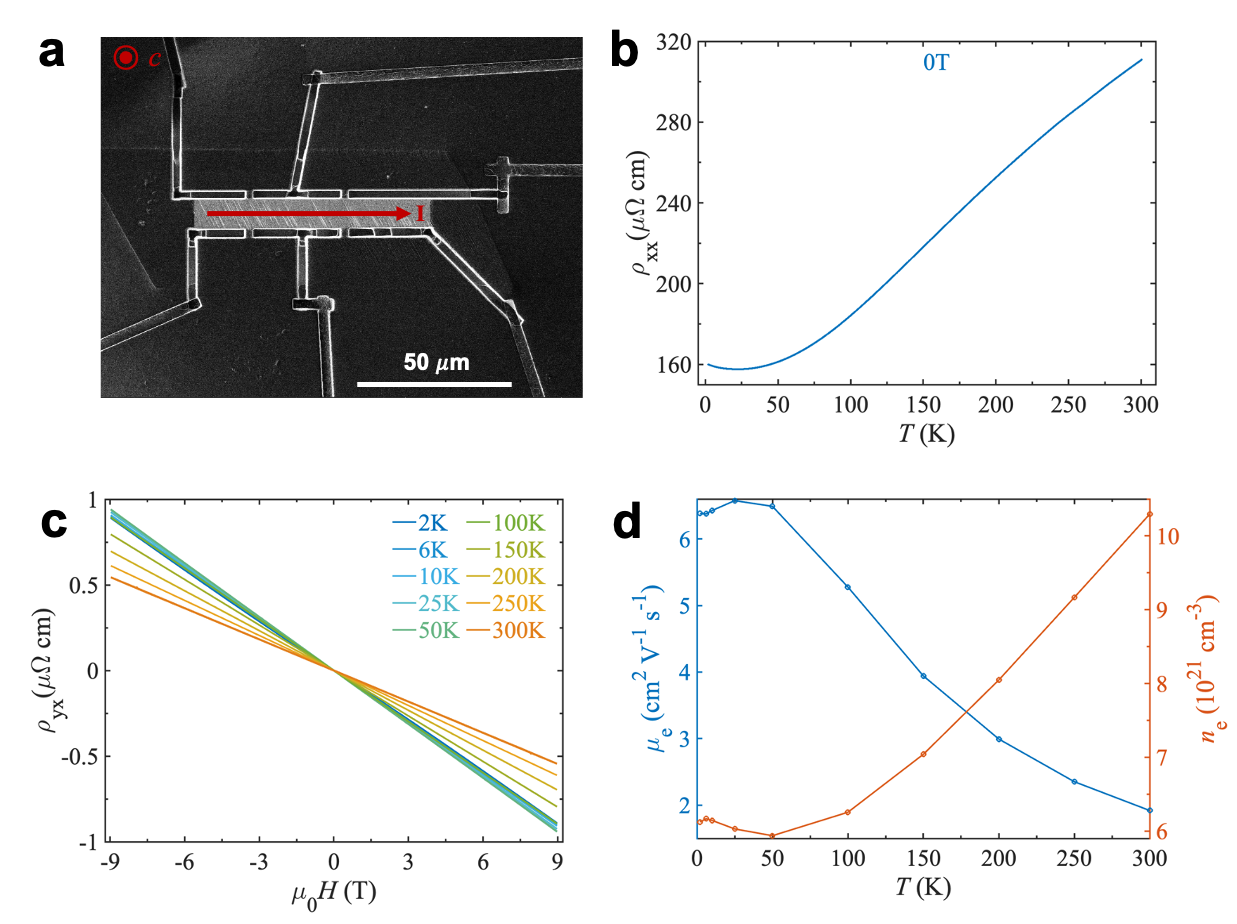}
\caption{\footnotesize{\label{FigS04}
(\textbf{\textsf{a}}) Transport device prepared using focused ion beam (FIB),
(\textbf{\textsf{b}}) Temperature dependent longitudinal resistivity at zero field
(\textbf{\textsf{c}}) Hall resistivity at different temperatures,
(\textbf{\textsf{d}}) mobility ($\mu_{\mathrm{e}}$) and carrier concentration ($n_{\mathrm{e}}$) at different temperatures. For all the transport, current is applied along ab direction and field along $c$.
}}
\end{figure}

\newpage
\section*{NMR Experiments - Spectra and site assignment}

\begin{figure}[h]
\centering
\includegraphics[width=.7\textwidth]{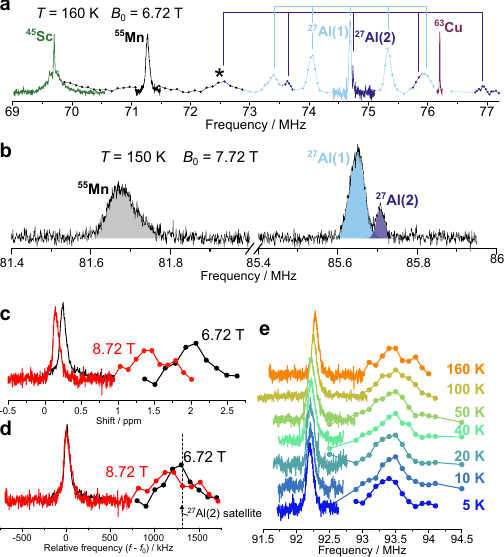}
\caption{\footnotesize{\label{FigS1}
(\textbf{\textsf{a}}) Combined multi\-nuclear spectrum at \SI{160}{K} and \SI{6.72}{T}. The $^{45}$Sc spectrum (green) was obtained from a broadband solid echo measurement, the black spectrum and data points denote selective spin echoes of the accessible $^{55}$Mn resonances, the two $^{27}$Al quadrupole patterns are shown in blue (selective spin echoes), and $^{63}$Cu from the rf-coil is shown in red. Signal intensities were adjusted for clarity.
(\textbf{\textsf{b}}) $^{55}$Mn (grey) and $^{27}$Al (blue) CTs as obtained from selective Spin echoes at \SI{150}{K} and \SI{7.72}{T}. 
(\textbf{\textsf{c}}) The Mn CT and its first high frequency satellite (labeled by $\ast$ in Fig.~1\textbf{a} of the main document) at \SI{160}{K} and for two fields, \SI{6.72}{} (black) and \SI{8.72}{T} (red). The CTs were obtained from single spin echoes while the satellites were received using narrow-band, frequency swept spin echoes to cover the full frequency range. The spectra are shown in units of ppm.
(\textbf{\textsf{d}}) The same spectra as in panel \textbf{b}, however, here presented in units of kHz. The vertical dashed line refers to the frequency of the second lower satellite of Al(2) (panel \textbf{a}).
(\textbf{\textsf{e}}) The temperature dependence of the Mn CT and its first high-frequency satellite for \SI{8.72}{T}.
}}
\end{figure}

\noindent
We begin with an overview of the NMR nuclei accessible in the system Sc$_3$Mn$_3$Al$_7$Si$_5$. Fig.~\ref{FigS1}\textbf{a} displays various spectra including the quadrupolar split spectrum of $^{45}$Sc (green), the $^{55}$Mn central transition (CT) next to one of its satellites (black), and two $^{27}$Al quadrupole patterns (light and dark blue). Also shown in red is the narrow resonance line of $^{63}$Cu from the rf-coil that serves as a shift reference. In the current report, we focus on the Mn resonances. Sc NMR is more difficult to measure as the compact quadrupole spectrum prevents well defined excitation conditions, while Al NMR was already investigated in the literature. 

The emphasis of the current report is laid in $^{55}$Mn$-$a quadrupole nucleus ($I=5/2$) of \SI{100}{\%} natural abundance$-$the central transition (CT) of which is shown in Fig. \ref{FigS1}{\textbf{b}} on the left, obtained using selective spin echoes at \SI{7.72}{T} and \SI{150}{K}. On the right, the two CTs of Al(1) and Al(2) ($I=5/2$ and NA$=\SI{100}{\%}$) are shown, reflecting the two chemically non-equivalent Al sites in the SMAS crystal structure. The resonances were normalized with respect to the stoichiometry, i.e., Mn was divided by 3, while the total Al spectrum was divided by 7. After correction for the transverse relaxation ($T_2$) the areas under the resonance lines between Mn and Al match perfectly (gray vs. blue), while the light blue Al(1) signal intensity is 6 times that of the dark blue Al(2). All in all, this evaluation confirms the expected crystal structure and the NMR to reflect a single crystalline sample.

The matching of the signal intensities of Mn and Al further proves the Mn spin system to be subject to a quadrupole interaction (similarly, nutation spectroscopy confirms the presence of a quadrupolar splitting of the Mn signal). As can be seen in panel \textbf{b} of Fig. \ref{Fig1}, the Al(1,2) lines are symmetrically split into 5 lines each, reflecting an in good approximation first order quadrupolar interaction. For Mn, the satellites are not as easily accessed. The peak labeled with ($\ast$) in panel \textbf{a} is one of them, another is hidden behind the $^{45}$Sc quadrupole pattern (green) around \SI{69.7}{MHz}. Their apparent broadening (as well as that of the Al satellites) reflect some degree of disorder and inhomogeneities in the system under investigation. DFT calculations find the quadrupole splitting frequencies (asymmetry) for Al(1) and Al(2) to be \SI{1.489}{MHz} ($\eta=0.64$) and \SI{1.063}{MHz} ($\eta=0$), respectively. For Mn, the DFT finds $^{\textrm{Mn}}\nu_{\mathrm{Q}}=\SI{5.266}{MHz}$ and $\eta=0$. The orientations of the corresponding electric field gradients (EFGs), however, only allow a direct comparison with experiment for the Al(2) spectrum, where the external field is aligned with $V_{ZZ}$. We find $^{\textrm{Al2}}\nu_{\textrm{Q,exp}}=\SI{1.085(50)}{MHz}$ in very good agreement with DFT. For Mn, the field dependent NMR shift as discussed in the next section allows an estimation of the underlying quadrupolar coupling, i.e., as from the second order quadrupole shift of the CT. We find $^{\textrm{Mn}}\nu_{\textrm{Q,exp}}=\SI{5.450(76)}{MHz}$, a value very close to the calculation, especially since the EFGs are not perfectly aligned with the external field.

The measurements shown in panel \textbf{c} and \textbf{d} provide further evidence that the peak labeled with ($\ast$) is not an additional NMR signal from an unknown source, because the distance to the Mn CT (narrow line) for different external fields would be the same in ppm units (panel \textbf{c}). A quadrupolar split system plotted in units of kHz, on the other hand, would be independent of the external field in first order, while slight deviations are expected from second order quadrupole interaction, which is what is found for the Mn CT and ($\ast$) (panel~\textbf{d}).

Tracing the Mn CT and one of its satellites ($\ast$) during cooling as shown in Fig.~\ref{FigS1}\textbf{e} confirms no changes in the local EFG at the Mn nuclei. Thus, temperature dependencies as related to the shift, line broadening, or relaxation (main manuscript) cannot be assigned to changes in the local charge symmetry.

\section*{The NMR shift and second order quadrupole contributions}

\begin{figure}[t]
\centering
\includegraphics[width=.7\textwidth]{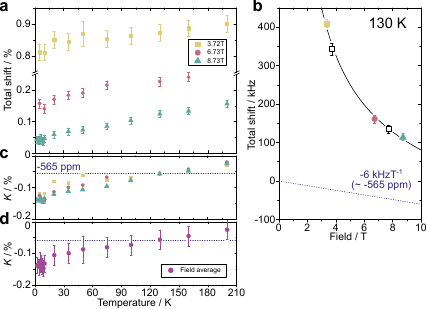}
\caption{\footnotesize{\textbf{The NMR shift and second order quadrupole contributions}\label{FigS2}
(\textbf{\textsf{a}}) The total Mn NMR shift as a function of temperature for fields of \SI{3.72}{}, \SI{6.73}{}, and \SI{8.73}{T} in units of ppm.
(\textbf{\textsf{b}}) The total MN shift as a function of the external field at \SI{130}{K}. The colored data points refer to panel \textbf{a}, the open black squares are obtained from other measurements. The black line represents a best fit to a function that relates the field dependence of the shift to a sum of a second order quadrupole term ($\propto B_0$) and a Knight shift term ($\propto B_0$). The resulting Knight shift (about \SI{-565}{ppm}) is given by the dotted line in the lower part of the graph.
(\textbf{\textsf{c}}) The total Mn shift for the 3 fields from panel \textbf{a} after subtraction of the corresponding second order quadrupole component. The dotted line refers to the extracted Knight shift component found in panel \textbf{b}.
(\textbf{\textsf{d}}) The field averaged shift from panel \textbf{c}.
}}
\end{figure}

\noindent
The NMR shift is strongly dependent on the external field. In particular, the shift in units of ppm grows steeply for decreasing magnetic fields as shown in Fig.~\ref{FigS2}\textbf{a}, proving it being governed by second order quadrupole interaction which carries inherently an inverse $B_0$ proportionality. We thus extracted the magnetic field dependence of the NMR shift at \SI{130}{K}, cf. panel \textbf{b}. We applied a fit $K_{\textrm{tot}}(B_0) = K\cdot B_0+K_{\textrm{Q}}(1/B_0)$. Here, $K$ is the wanted Knight shift and \cite{Freude1993} 

\begin{equation}\label{KQ}
K_{\textrm{Q}}=-\frac{\nu_{\textrm{Q}}^2}{6\gamma B_0}\left\lbrace I(I+1)-\frac{3}{4}\right\rbrace\left(-\frac{3}{8}\right)
\end{equation}

\noindent
the second order quadrupole shift of the CT assuming a symmetric tensor ($\eta=0$) with the external field perpendicular to the main principle axis. The fit yielded the aforementioned $^{\textrm{Mn}}\nu_{\textrm{Q,exp}}=\SI{5.450(76)}{MHz}$. Note, the 2nd order shift can only be an approximation because the tensor symmetry is suggested by DFT, not measured, and the actual orientation of the EFG is unknown while Eq.~\ref{KQ} assumes $V_{ZZ}$ to lie in the kagome plane, i.e., to be perpendicular to $B_0$. DFT calculations find the EFG's main axis to be slightly tilted out of the $ab$-plane (i.e., not alinged with a crystal axis). Most importantly, however, the uncertainty only affects the resulting value of the underlying quadrupole interaction \emph{but not} $K$ which is what we are looking for. The fits yields $K=-\SI{6}{kHzT^{-1}}$ or \SI{-565}{ppm} (blue dotted line in Fig.~\ref{FigS2}\textbf{b}). We next subtracted the quadrupole shift for each field from the respective total shift data. The results are given in panel c. As it is already apparent from the total shift shown in panel \textbf{a}, the overall temperature dependence in ppm is very similar for all 3 fields, and we thus averaged the quadrupole-corrected shift data to obtain the final shift $T$ dependence as shown in Fig.~\ref{FigS2}\textbf{d}. The shift is found to be negative, and its temperature dependence to cover about \SI{1000}{ppm} during cooling from about \SI{-300}{ppm} at \SI{200}{K} to $\sim\SI{-1300}{ppm}$ at \SI{2.1}{K}. Note that the rather large error bars in the graph refer to uncertainties in the vertical position of the set of shift data points rather than to uncertainties of individual shift values, i.e., the temperature dependence of the shift is more reliable while there can be some collective offset.

\section*{Spin-lattice relaxation}

\begin{figure}[t]
\centering
\includegraphics[width=.7\textwidth]{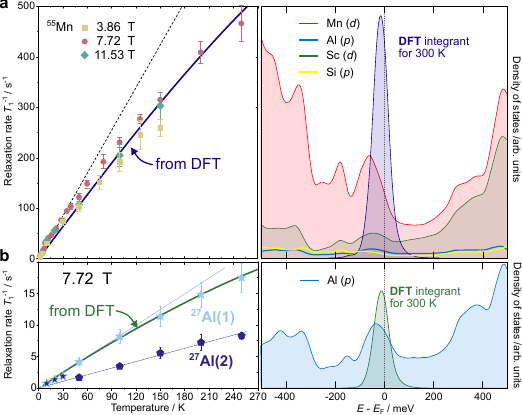}
\caption{\footnotesize{\label{FigS3}
(\textbf{\textsf{a}}) The Mn spin-lattice relaxation rate $T_1^{-1}$ as a function of temperature for 3 external fields. The dashed line represents a Korringa behavior with $T_1^{-1}\propto T$ while the blue solid line gives what follows from the actual density of states (DOS). In the right panel the density of states (DOS) for the 4 largest contributions at the Fermi level is shown, including the integrant of Eq.~\eqref{T1} for \SI{300}{K} (dark blue). 
(\textbf{\textsf{b}}) $T_1^{-1}(T)$ of the two sites Al(1,2) for \SI{7.72}{T} beside the Al $p$ DOS.
}}
\end{figure}

\noindent
In Fig.~\ref{FigS3}\textbf{a} we show the temperature dependent spin-lattice relaxation rate $T_1^{-1}$ of the Mn CT for three magnetic fields. The values were obtained from the recovery of the selectively saturated CT assuming a purely magnetic relaxation mechanism. That is, we applied the following fit function to the evolution of the CT signal intensity ($M(t)$ from Hahn-echoes) as a function of the relaxation delay:

\begin{equation}\label{T1}
M(t)=M_0\left[1-f\cdot\biggl\{\frac{1}{35}\exp\left(-\frac{t}{T_1}\right)+\frac{8}{45}\exp\left(-\frac{6t}{T_1}\right)+\frac{50}{63}\exp\left(-\frac{15t}{T_1}\right)\biggl\}\right] .\\[0.2cm]
\end{equation}

\noindent
Here, $M_0$ is the equilibrium signal intensity (full magnetization), $f$ the inversion factor ($f=1$ for saturation recovery measurements), and $T_1$ the spin system's spin lattice relaxation time. To evaluate the individual CT lifetime ($T_{\mathrm{1,exp}}$) as related to the high temperature spin-spin relaxation in Fig.~\ref{Fig2}\textbf{a} (gray open symbols) we approximated the recovery after saturation with a single exponential:

\begin{equation}\label{T1exp}
M(t)=M_0\left[1-f\cdot\biggl\{\exp\left(-\frac{t}{T_{\mathrm{1,exp}}}\right)\biggl\}\right] .\\[0.2cm]
\end{equation}

\noindent
As expected, the CT lifetime is about one order of magnitude shorter than the characteristic lifetime of the spin system, while the temperature dependence is in very good approximation the same.

The relaxation rate as shown in Fig.~\ref{FigS3}\textbf{a} grows sub-linearly with temperature (compare the dashed line representing a metallic system with $T_1^{-1}\propto T$), while there is no systematic change with the applied field. Based on the Mn $d$ density of states (DOS) as shown in the right panel of Fig.~\ref{FigS3}\textbf{a} (red) and 
\begin{equation}\label{T1dos}
\frac{1}{T_1}\propto \int f(E-\mu_c)[1-f(E-\mu_c)]D(E)^2\mathrm{d}E\ ,
\end{equation}
where $f(E-\mu_c)$ and $D(E)$ denote the Fermi function and the energy dependence of the DOS, respectively, we calculated the expected temperature dependence of the spin-lattice relaxation (blue solid line in Fig.~\ref{FigS3}\textbf{a}). The integrant for \SI{300}{K} is shown exemplarily in the right panel. As can be seen, the peak in the DOS as related to bands with vanishing dispersion is too far to affect the Mn $T_1$ severely. The curvature around $E_{\mathrm{F}}$, however, yields a sub-linear behavior for higher temperatures, quite similar to the measurements. Deviations at lower temperatures become visible that lead to the peak in $(T_1T)^{-1}$ around \SI{10}{K} as discussed in the main manuscript.

Below, in panel \textbf{b}, the spin-lattice relaxation rates of Al(1) and (2) for selected temperatures are shown next to the Al $p$-orbital DOS. Due to excessive line broadening, below \SI{60}{K} the two lines can no longer be separated. $T_1$ values were also obtained from selective saturation recovery measurements and Eq.~\ref{T1}. The temperature dependencies seem also to be slightly curved toward high temperature, which is, again, captured in good approximation by the DOS approach (green solid line). As expected from the much smaller Fermi level DOS (cf. Fig.~\ref{Fig3}\textbf{a}), Al relaxation is much slower than Mn relaxation, about a factor of 25. Between Al(1) and Al(2) we find a difference in the relaxation by about a factor of 2.

\section*{Low temperature spin-spin relaxation}

\begin{figure}[t]
\centering
\includegraphics[width=\textwidth]{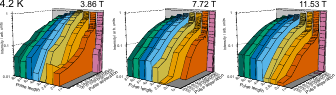}
\caption{\footnotesize{\textbf{}\label{FigS4}
Pulse length dependent spin-echo decay at \SI{4.2}{K} for 3 different external fields, \SI{3.86}{T} (\textit{left}), \SI{7.72}{T} (\textit{middle}), and \SI{11.53}{T} (\textit{right}). Each wall represents the signal intensity (area under the resonance line in Fourier space) evolution for increasing pulse separation times. Using a sum of a single exponential ($T_2$) and a Gaussian ($T_{\textrm{2G}}$) decay, we extracted information about the change in $T_2/T_{\textrm{2G}}$ as well as the corresponding initial intensities (for $t=0$). The resulting data set is provided in Fig.~\ref{Fig3}\textbf{c} and \textbf{d} of the main manuscript.
}}
\end{figure}

\noindent
Transverse relaxation measurements at \SI{2.1}{K} and \SI{4.2}{K} revealed that a fraction of nuclei decays very slowly beside other nuclei with the expected rapid relaxation (cf. Fig.~\ref{Fig3}\textbf{a} and \textbf{b} blue circles in the main manuscript). We conducted a detailed investigation at \SI{4.2}{K} to quantify this behavior. The results in terms of the echo-decay envelops$-$signal intensity as a function of the pulse separation time $\tau-$for various excitation band widths (excited spin fractions) and for 3 external fields is shown in Fig.~\ref{FigS4}. 

\begin{figure}[t]
\centering
\includegraphics[width=.7\textwidth]{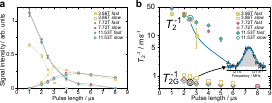}
\caption{\footnotesize{\textbf{}\label{FigS5}
(\textbf{\textsf{a}}) The relative signal intensities of the two relaxation components from extrapolating the spin-echo decays to $\tau\rightarrow0$. As the pulse length increases and the number of excited spins (total intensity) decreases, the \emph{slow} component gradually replaces the \emph{fast} component.
(\textbf{\textsf{b}}) The spin-spin relaxation rates of the two components (\emph{fast} and \emph{slow}) for each excitation condition (pulse length) as extracted from the measurements in Fig.~\ref{FigS4}. The inset shows the spectrum for a Hahn-echo at \SI{11.53}{T} for a very short and a long pulse separation time. The spectra are nearly identical in resonance frequency and width (as well as in their spin-lattice relaxation), proving that the involved spin sit in the same chemical environment. 
}}
\end{figure}

Two relaxation regimes can be observed. For the full excitation of the resonance line ($\SI{1}{\mu s}$), the entire resonance relaxes extremely fast, as expected from higher temperatures (Fig.~\ref{Fig2}\textbf{b}). Once the excitation fraction is reduced, a slow component emerges below the fast component. As the number of excited nuclei is further reduced, the fraction of \emph{slow} nuclei grows while the fraction of \emph{fast} nuclei reduces, until eventually the entire NMR signal decays slowly and the fast component is replaced. Both regimes differ widely not only in terms of the relaxation time, but also in their echo decay function. The fast component$-$and this holds also for any other measurement at higher temperature$-$decays exponentially ($T_2$), while the slow component features a pronounced Gaussian decay ($T_{\textrm{2G}}$). The corresponding initial signal intensities and relaxation rates are shown in Fig.~\ref{FigS5}. The fast component rapidly slows as observed for the excitation width dependent relaxation at higher temperatures (Fig.~\ref{Fig2}\textbf{e}). The slow component also slows down as the excitation width is reduced, however, not as fast as the fast component. The weaker dependence on the excitation fraction and the time scale strongly suggest direct dipole-dipole interaction to cause the observed slow decay. As the inset in panel \textbf{b} shows, the slow component covers the same frequency range as the fast component, while spin-lattice relaxation measurements prove both components are subject to the same electronic environment also in terms of dynamics. That is, the two regimes represent the \emph{same} nuclei and there is no phase segregation.

\section*{Data availability}
\noindent
The data that support the ﬁndings of this study are available from the corresponding
authors upon request.

\section*{Code availability}
\noindent
The codes that support the ﬁndings of this study are available from the corresponding
authors upon request.

\section*{Competing interest}
\noindent
The authors declare no competing interests.

\end{document}